\documentclass[12pt]{iopart}
\usepackage[latin1]{inputenc}
\usepackage[T1]{fontenc}
\usepackage[english]{babel}
\begin{document}
\title[Isotropisation of the minimally coupled scalar-tensor theory...]{Isotropisation of the minimally coupled scalar-tensor theory with a massive scalar field and a perfect fluid in the Bianchi type $I$ model}
\author{Stéphane Fay\\
14 rue de l'Etoile\\
75017 Paris\\
France\\
\small{Steph.Fay@Wanadoo.fr}}
\begin{abstract}
We look for necessary conditions such that minimally coupled scalar-tensor theory with a massive scalar field and a perfect fluid in the Bianchi type $I$ model isotropises. Then we derive the dynamical asymptotical properties of the Universe.
\\
\\
\\
Published in Classical and Quantum Gravity copyright 2002 IOP Publishing Ltd\\
Classical and Quantum Gravity, Vol 19, 2, 2002.\\
http://www.iop.org
\end{abstract}
\pacs{11.10.Ef, 04.50.+h, 98.80.Hw, 98.80.Cq}
\maketitle
\section{Introduction}\label{s0}
In this work, we will study the isotropisation of a minimally coupled scalar tensor theory with a massive scalar field and a perfect fluid in the Bianchi type $I$ model.\\
Lets give some reasons about our choice for this geometrical framework. Although our present Universe seems in agreement with the cosmological principle, it could be necessary to partly release it for the early times if we want to explain the formation of large-scale structures. Moreover, the observations of the cosmic microwave background by COBE show some small inhomogeneities. These observational facts lead us to assume a more general geometry than this of the isotropic models i.e. the Friedman-Lemaitre-Robertson-Walker (FLRW) ones. The simplest generalisation is to leave the isotropy hypothesis and to consider the Bianchi models. About the nine Bianchi models, some of them accept FLRW models as exact solutions. Hence, Bianchi type $I$ models contain the flat FLRW ones , the Bianchi type $V$ ones contain the open FLRW ones, and the Bianchi type $IX$, the closed ones. At the present time, despite some more and more powerful observational tools, we do not know in which type of Universe we live. However, Boomerang experiment favours a flat Universe\cite{Lan00} and the same conclusion could be drawn from the presence of inflation\cite{Per99,Rie98} although this phenomenon has to be confirmed by deepest observations and could be compatible with other types of geometries\cite{ChiYam99}. All these facts justify the interest of the Bianchi type $I$ model.\\

As a physical framework, we have chosen to study a scalar-tensor theory minimally coupled to a massive scalar field $\phi$. The geometrical part of its Lagrangian writes as this of the General Relativity, which describes with a high precision the local dynamics of our Universe for the weak fields. Taking into accounts one or several scalar fields could be one of the key for a theory able to explain the physics of the early times. They are predicted by unification theories whose low energy limit could be a scalar-tensor theory. They also appear during dimensional reduction of Kaluza-Klein type theories. In  addition, we will also assume that $\phi$ is a massive field. The reason is that inflation could be the consequence of the presence of a cosmological constant whose currently observed value and particle physics predicted value differ from 120 orders of magnitude: this is the so called cosmological problem. To explain this huge discrepancy, a solution could be to consider that the cosmological constant is in fact a variable potential $U$ representing the coupling of $\phi$ with itself. Moreover, we will consider a Brans-Dicke coupling function $\omega$ between the scalar field and the metric. The theory thus described has been studied in \cite{Fay01}. Here, we will generalise it by adding a perfect fluid. Associating a scalar field and a perfect fluid could be a way to explain the nature of dark matter, if the first one plays the same dynamical role as the second one as suggested by the quintessence or tracking models\cite{BarPie99, MatGuzUre99}. In the quintessence model, the scalar field slowly rolling down its potential such that the ratio of its pressure and energy density, w, be a constant belonging to the range $\left[-1,0\right]$. One problem of the quintessence model is the cosmic coincidence problem: why the present energy density of the scalar field would be of the same order as this of the matter energy density. One possible solution to this question could be to consider a special form of quintessence, called tracker model, for which $w$ is time varying, and works like an attractor solution. Thus, late times cosmology would be independent of the early conditions.\\

From an observational point of view, the standard model that seems to emerge today is the same as this described above with $\omega=0$ and $U=cte$. However, this particular one is far from being satisfactory (as instance, it does not solve the cosmological problem). Hence, more general theories leaving $\omega$ and $U$ undetermined have to be studied and some efforts are currently done to try and guess what could be some limitation on their forms and values. As instance, in \cite{BoiEspPolSta00}, it is shown how from the observations one can determine the Lagrangian of a scalar tensor theory. In \cite{SerAli96}, it is demonstrated that scalar tensor theories can be compatible both with primordial nucleosynthesis and solar-system experiments with cosmological models very different from the FLRW ones. In \cite{SchWil01}, it is shown how new bounds on $\omega$ could be derived from future space gravitational wave interferometers, thus allowing to test scalar tensor gravity. All these works are related to observational cosmologies and aim to derive some satisfactory limits on $\omega$ and $U$ as we try to do it from a theoretical ground in the present paper.\\

Mathematically we will study the isotropisation of the Universe in the same way as in \cite{Fay01}, i.e.  by associating the Hamiltonian formalism of Arnowitt, Deser and Misner (ADM)\cite{Nar72, MatRyaTot73} which allows getting first order dynamical system equations with the dynamical system methods\cite{WaiEll97}. Our goal will be to determine the necessary conditions for the isotropisation of the theory described above and the asymptotical dynamical behaviours of the Universe at late times. The plan of this work is the following. In the second section, we establish the field equations of the Hamiltonian formalism. In the third section, we analyse their dynamics. In the fourth section, we discuss the physical meaning of our results.
\section{Field equations}\label{s1}
In this section, we will calculate the field equations of the minimally coupled scalar-tensor theory with a massive scalar field and a perfect fluid. The action is written:
\begin{equation} \label{1}
S=(16\pi)^{-1}\int \left[R-(3/2+\omega(\phi))\phi^{,\mu}\phi_{,\mu}\phi^2 -U(\phi)+16\pi c^4 L_m\right]\sqrt{-g}d^4 x
\end{equation}
with $\phi$ the scalar field, $\omega$ the coupling between the scalar field and the metric, $U$ the potential and $L_m$ the Lagrangian density of the matter. We will consider a perfect fluid with an equation of state $p=(\gamma-1)\rho$, $p$ and $\rho$ being respectively the pressure and the density of the fluid. For $\gamma=0$, $\gamma=1$ and $\gamma=4/3$ we get respectively the equation of state describing the vacuum energy, a dust and radiation fluid. As the first case can be assimilated to the presence of a cosmological constant already studied in \cite{Fay01}, we will not consider it and will assume that $\gamma\in\left[1,2\right]$. We will use the following form of the metric:
\begin{equation}
ds^2 = -(N^2 -N_i N^i )d\Omega^2 + 2N_i d\Omega\omega^i + R_0 ^2 g_{ij}\omega^i \omega^j 
\end{equation}
The $\omega^i$ are the 1-forms defining the homogeneous space of the Bianchi type $I$ model and the $g_{ij}$ are the metric functions. To derive the expression of the ADM Hamiltonian, we proceed in the same way as \cite{Nar72} or \cite{Fay00A}. We rewrite the metric functions by using the Misner parameterisation\cite{Mis69}:
\begin{eqnarray*}
g_{11}&=&e^{-2\Omega+\beta_++\sqrt{3}\beta_-}\\
g_{22}&=&e^{-2\Omega+\beta_+-\sqrt{3}\beta_-}\\
g_{33}&=&e^{-2\Omega-2\beta_+}\\
\end{eqnarray*}
The $\beta_\pm$ functions describe the anisotropy of the Universe and $\Omega$ its isotropic part. We can derive the field and the Klein-Gordon equations of the Lagrangian formulation by varying the action with respect to the metric functions and the scalar field. Then using the Bianchi identities and the Klein-Gordon equation, we can find the conservation law for the energy impulsion of the perfect fluid, $T^{0\mu}_{;\mu}=0$, and thus the well known relation between the energy density and the 3-volume $V$ of the Universe: $\rho=\rho_0 V^{-\gamma}$, $\rho_0$ being an integration constant and $V=(\prod_i g_{ii})^{1/2}=e^{-3\Omega}$. To find the Hamiltonian of the ADM formalism, we have to express the action with the variables $\beta_+$, $\beta_-$ and their conjugate momentum $p_+$ and $p_-$. Then, we vary it with respect to $N$ which plays the role of a Lagrange multiplier. We get a constraint equation from which we can derive the Hamiltonian. The reader interested by full details of its derivation when no perfect fluid is present can find it in \cite{Fay01}. Finally, we get:
\begin{equation} \label{4}
H^2 = p_+ ^2 +p_- ^2 +12\frac{p_\phi ^2 \phi^2}{3+2\omega}+24\pi^2 R_0 ^6 e^{-6\Omega}U+\delta e^{3(\gamma-2)\Omega}
\end{equation}
$p_\phi$ is the conjugate momentum of the scalar field and $\delta$ a constant equal to $(\gamma-1)\rho_0$. It is a positive constant when $\gamma\in\left[1,2\right]$ and the energy density of the perfect  fluid is positive. We will assume it is the case. From (\ref{4}), we derive the Hamilton's equations:
\begin{equation} \label{5}
\dot{\beta}_ \pm = \frac{\partial H}{\partial p_ \pm}=\frac{p_\pm}{H}
\end{equation}
\begin{equation} \label{6}
\dot{\phi}=\frac{\partial H}{\partial p_\phi}=\frac{12\phi^2 p_\phi }{(3+2\omega)H}
\end{equation}
\begin{equation} \label{7}
\dot{p}_\pm=-\frac{\partial H}{\partial \beta_ \pm}=0
\end{equation}
\begin{equation} \label{8}
\dot{p}_\phi=-\frac{\partial H}{\partial \phi}=-12\frac{\phi p_\phi ^2}{(3+2\omega)H}+12\frac{\omega_\phi \phi^2 p_\phi ^2 }{(3+2\omega)^2 H}-12\pi^2 R_0 ^6 \frac{e^{-6\Omega}U_\phi }{H}
\end{equation}
\begin{equation} \label{9}
\dot{H}=\frac{dH}{d\Omega}=\frac{\partial H}{\partial \Omega}=-72\pi^2 R_0 ^6 \frac{e^{-6\Omega}U}{H}+3/2\delta(\gamma-2)\frac{e^{3(\gamma-2)\Omega}}{H}
\end{equation}
A dot means a derivative with respect to $\Omega$. If we compare them with these got when no perfect fluid is present\cite{Fay01}, we remark that only the equation (\ref{9}) is modified. We will choose $N^i=0$ and then we calculate that $N$ \cite{Nar72}\cite{MatRyaTot73} can be written\footnote{In few words, this result can be recovered in the following way. We first rewrite the action under a Hamiltonian form, i.e. $S=\int (g_{ij}\frac{\partial \Pi_{ij}}{\partial t}-NH-N_iH_i)d^4x$, $\Pi_{ij}$ being the conjugate momentum of the metric functions $g_{ij}$, $N$ and $N_i$ the lapse and shift functions, $H$ and $H_i$ the super Hamiltonian and super momentum. Then, by varying the expression thus obtained with respect to $\Pi_{ij}$, we derive an expression for $\partial g_{ij}/\partial t$. Developing $-1/2(-g)^{-1/2}\delta g/\delta\Omega$ as a functions of the $g_{ij}$ and using the above mentioned expression, we find (\ref{9a}).}:
\begin{equation} \label{9a}
N=\frac{12\pi R_0^3e^{-3\Omega}}{H}
\end{equation}
The relation between the $\Omega$ time and the proper time $t$ being $dt=-Nd\Omega$, we deduce that $t$ is a decreasing function of $\Omega$ for all positive Hamiltonian.
\section{Isotropisation conditions and asymptotical behaviours}\label{s2}
In the first subsection, we rewrite the field equations with new normalised variables. In the second one, we study mathematically the system thus obtained.
\subsection{Rewriting of the field equations with normalised variables}\label{s21}
The equations (\ref{4}-\ref{9}) form a first order system that we wish to rewrite with the following variables:
\begin{equation} \label{10a}
x=H^{-1}
\end{equation}
\begin{equation} \label{10b}
y= e^{-3\Omega}\sqrt{U}H^{-1}
\end{equation}
\begin{equation} \label{10c}
z=p_{\phi}\phi(3+2\omega)^{-1/2}H^{-1}
\end{equation}
They are independent each other's since the first one depends on $H$, the second one on $H$ and $\phi$ and the third one on $H$, $\phi$ and $p_\phi$. The forms of $y$ and $z$ show that the potentials $U$ have to be positive and the Brans-Dicke coupling function $\omega$ must be larger than $-3/2$ so that the variables be real. These are usual hypothesis in cosmology. The constraint (\ref{4}) is then written:
\begin{equation} \label{11}
p^2x^2+R^2y^2+12z^2+k^2=1
\end{equation}
where we have put to simplify the equations $k^2=\delta e^{3(\gamma-2)\Omega}H^{-2}=\delta x^\gamma y^{2-\gamma}U^{\gamma/2-1}$. The constants $p$ and $R$ are defined by $p^2=p_+^2+p_-^2$ and $R^2=24\pi^2R_0^6$. Rewriting the equations (\ref{6}), (\ref{8}) and (\ref{9}), we get:
\begin{equation} \label{12}
\dot{x}=3R^2y^2x-3/2(\gamma-2)k^2x
\end{equation}
\begin{equation} \label{13}
\dot{y}=y(6\ell z+3R^2y^2-3)-3/2(\gamma-2)k^2y
\end{equation}
\begin{equation} \label{14}
\dot{z}=y^2(3Rz-R^2/2\ell)-3/2(\gamma-2)k^2z
\end{equation}
with $\ell=\phi U_\phi U^{-1} (3+2\omega)^{-1/2}$. From the first one, we deduce that $x$ is a monotonic function of constant sign and then that no homoclinic orbit are allowed.\\
In the \ref{Ap1}, we show that even if $\delta<0$, i.e. if we consider negative energy density, the equilibrium is not compatible with the divergence of $k$. 
\subsection{Mathematical study of the first order system equations}\label{s22}
In the first subsection, we examine the values of $x$ and $k$ allowing the isotropisation. In the second one, we look for the equilibrium points corresponding to a stable isotropic state for the Universe.
\subsubsection{Values of $x$ and $k$ compatible with the isotropisation}\label{s221}
Assuming that late times correspond to the divergence of the proper time $t$, an isotropic and stable state is such that $\beta_\pm$ tend toward some constants with $d\beta_\pm/dt\rightarrow 0$. By using the expression for the lapse function, we calculate that it happens only when $\Omega\rightarrow -\infty$. The Universe is then expanding. It corresponds to late times epoch when the Hamiltonian is positive and justify our assumption on the asymptotical value of $t$ since there is no physical reason such that the diverging expansion of the Universe take place for a finite value of the proper time. Since in the same time, $\beta_\pm$ should tend toward some constants, we should also have $\dot{\beta}_\pm\rightarrow 0$, i.e. $x\rightarrow 0$. Consequently, a stable isotropic state can be reached only in the plane $x=0$ of the phase space when $\Omega$ diverges negatively. We will have to check if these two conditions do not excluded each others.\\
\\
What about the value of $k$? From the constraint, we see that $k^2\leq 1$. Then, by considering the expression $k^2=\delta H^{-2}e^{3(\gamma-2)\Omega}$ and the equation for $\dot{H}H$ issued from (\ref{9}), we deduce that:
\begin{itemize}
\item $k^2$ will tend to vanish when $U>V^{-\gamma}$.
\item $k^2$ will tend toward a constant different from zero when $x=0$ if the Hamiltonian tends toward $H=H_0e^{3/2(\gamma-2)\Omega}$. Then $k^2\rightarrow\delta H_0^{-2}$ and $U\propto V^{-\gamma}$.
\item $k^2$ will tend toward $1$ for any potential such that $U<V^{-\gamma}$. 
\end{itemize}
The first case corresponds to an asymptotically dominated scalar field Universe. It has already been studied in \cite{Fay01}. It will concern any potential tending toward a constant or diverging since the 3-volume will diverge when isotropisation arises and then $V^{-\gamma}\rightarrow 0$.

The second case corresponds to a potential behaving asymptotically as the energy density of the perfect fluid, $U\propto\rho$. It is different from trackers solutions which are such that asymptotically $\rho_\phi\propto\rho$, $\rho_\phi$ being the scalar field energy density: then, it is the scalar field energy density that mimics this of the perfect fluid and $U\leq \rho$. However, in both cases, the metric functions behave asymptotically as the Universe was filled only with a perfect fluid whose density is increased by the presence of the scalar field. Hence, when $U\propto V^{-\gamma}$, dark matter and coincidence problems could be explained in the same way as trackers solutions try to do it. For this reason, we will name these solutions "trackers like" solutions.

The third case corresponds to a Universe dominated by the perfect fluid. Effectively, if we use a Lagrangian formulation to rewrite the field equations, we remark that when $U<V^{-\gamma}$, in the space field equations we can neglect the potential regarding the term of the perfect fluid. Thus, the asymptotical solutions of the metric function will take the same form as these of a theory without scalar field. However, in the constraint equation, the scalar field is not negligible.  We will see below that this case always implies that $\ell$ diverges, that is also possible in the previous case. 

In the following subsection, only the non asymptotically scalar field dominated cases that contain the trackers solutions, will be studied, the results of the scalar field dominated case being identical to these of \cite{Fay01}.
\subsubsection{Isotropic stable equilibrium states for non asymptotically scalar field dominated Universe}\label{s222}
We find four equilibrium points in the plane $x=0$, with $k\not =0$ and respecting the constraint if $k^2=1-3\gamma(2\ell^2)^{-1}$. This last condition is not a fine-tuning. In fact, the expression for $k$ contains the integration constant $\delta$ which is hence determined by the constraint equation. As evoked above, it shows that when $\ell\rightarrow \infty$, $k\rightarrow 1$. Moreover, since $k^2$ and $\delta$ are positive, we derive that $\ell^2\not\in\left[0,3/2\gamma\right]$ and thus eliminate two of the equilibrium points which are not real under this condition. The two remaining ones are then defined by $(y,z)=(\pm(2R\ell)^{-1}\sqrt{3\gamma(2-\gamma)},(4\ell)^{-1}\gamma)$. They are real as long as $\gamma<2$. The equilibrium states taking place in $\Omega\rightarrow -\infty$, they are sources for the $\Omega$ time and sinks for the proper times $t$ when the Hamiltonian, which can be assimilated to an energy, is positive. Linearising the equation (\ref{12}) in the neighbourhood of the equilibrium points allows us to calculate $x(\Omega)$ and, using the relation $dt=-Nd\Omega$, we get that asymptotically $e^{-\Omega}$ tends toward $t^{\frac{2}{3}\gamma^{-1}}$ whatever $\ell$. It follows that if the potential behaves like the energy density of the perfect fluid, near the equilibrium, both tend toward zero as $t^{-2}$. We have also checked that $x(\Omega)\rightarrow 0$ when $\Omega\rightarrow -\infty$, thus showing the compatibility of these two limits necessary for isotropisation.\\
In the next section, we discuss about physical meaning of these results, compare them with other papers and make some applications.
\section{Discussion}\label{s3}
In this paper, we have considered the isotropisation of a scalar-tensor theory minimally coupled to a massive scalar field with a perfect fluid for the Bianchi type $I$ model. The following discussion is divided in three parts. The first one contains the set of results of this work, the second one some comparisons with other papers and the third one, some applications concerning the most studied forms of potentials.\\	
\\
We have seen that three cases can be distinguished depending on the fact that the potential is larger, smaller or behaves in the same way as the energy density of the perfect fluid. For the first case, we recall the result we have got in \cite{Fay01}, adapting its terms to the present paper:\\
\\
\emph{\underline{Asymptotically dominated scalar field Universe}\\When the potential is asymptotically larger than the energy density of the perfect fluid, a necessary condition for isotropisation of the minimally coupled scalar field with a massive scalar field $\phi$ and a perfect fluid, whatever the Brans-Dicke coupling function $\omega$ and the potential $U$ considered, will be that $\phi U_\phi U^{-1} (3+2\omega)^{-1/2}$ tends toward a constant $\ell$ with $\ell^2<3$. It arises at late times if the Hamiltonian is positive, at early times otherwise. If $\ell\not =0$ the metric functions tend toward $t^{\ell^{-2}}$. The Universe is expanding and will be inflationary if $\ell^{2}<1$. If $\ell=0$, it tends toward a De Sitter model.}\\
\\
It includes all the diverging potentials or these tending toward a cosmological constant at late times as it could be the case for our present Universe. The first new result of this study concerns the potentials that mimic asymptotically the energy density $\rho$ of the perfect fluid. In this case, we can consider that their effect is equivalent to increase $\rho$.\\
\\
\emph{\underline{Isotropisation of trackers like theories:}\\When the potential asymptotically behaves like the positive energy density of the perfect fluid with an equation of state $p=(\gamma-1)\rho$ and $\gamma\geq 1$, necessary conditions for isotropisation will be that $\phi^2 U_\phi^2 U^{-2} (3+2\omega)^{-1}$ tends toward a constant $\ell^2$ larger that $\frac{3}{2}\gamma$ and $\gamma<2$. The isotropisation always arises at late(early) times when the Hamiltonian is positive(negative). Then the metric functions tend toward the attractor $t^{\frac{2}{3}\gamma^{-1}}$, the Universe is expanding, non-inflationary and the potential and density $\rho$ tend toward zero as $t^{-2}$.}\\
\\
Finally, the last case concerns a theory asymptotically dominated by the matter. Then, the potential $U$ is negligible regarding the energy density $\rho$:\\
\\
\emph{\underline{Asymptotically matter dominated Universe}:\\
When the potential is asymptotically smaller than the energy density of the perfect fluid, the necessary conditions for isotropisation are the same as for trackers like theories but the quantity $\phi^2 U_\phi^2 U^{-2} (3+2\omega)^{-1}$ is always diverging.}\\
\\
To complete these two last results, we need to determinate the asymptotical value of $\phi$ near the equilibrium. From it, we will be able to determine the asymptotical value of the potential relative to $V^{-\gamma}$ and the constant $\ell$. Using the equation (\ref{6}) and writing it near the equilibrium, we deduce that:\\
\\
\emph{\underline{Asymptotical behaviour of the scalar field near the equilibrium}:\\
The asymptotical behaviour of the scalar field near the equilibrium state when it does not dominate the Universe is this of the function $\phi$ defined by the differential equation $\dot{\phi}=3\gamma UU_\phi^{-1}$ when $\Omega\rightarrow -\infty$.}\\
\\
Note that in \cite{Fay01}, it has been shown that the corresponding equation for the scalar field when it dominates the Universe is $\dot{\phi}=2\phi^2U_\phi(3+2\omega)^{-1}U^{-1}$. All these results are independent from an initial state for the Universe except the sign of $H$ which have to be initially positive so that the isotropisation take place at late times. Remark also that they can not be applied to a theory without potential since then the variable $y$ and thus $\ell$ is not defined.\\
\\

In this second part, we compare our results related to the non-asymptotically dominated scalar field Universe with these of other papers.\\
Hence, in \cite{HebWet00} is studied what is called the "Quintessential adjustment of the cosmological constant". The quintessence phenomenon is considered for an FLRW model and leads naturally to a vanishing potential. It is what we observe here, the quintessential solutions being such that $U\leq V^{-\gamma}$, since then necessary conditions for isotropisation imply that \emph{the potential tends toward zero at the most as $t^{-2}$}. It could thus solve the cosmological constant problem.\\
In \cite{ChiJakPav00} where the General Relativity with a perfect fluid and a quintessential matter is studied for a flat isotropic model, it has been demonstrated that \emph{the solving of the coincidence problem was not compatible with inflation}. This result is here generalised to any isotropising Bianchi type $I$ model whatever $\omega$ and $U$ since for inflation being present at late time we would need that $\gamma<2/3$, which is not the case for ordinary matter.\\
In \cite{SanKalWag98}, it has been shown that scalar-tensor theory with a potential and a perfect fluid can \emph{have as late time attractor the General Relativity}. It corresponds to what we have found in this work since when necessary conditions for isotropy are respected the metric functions asymptotically tend toward $t^{\frac{2}{3}\gamma^{-1}}$. Hence, when $\gamma=1$, we have a dust fluid and the Universe tends toward an Einstein De-Sitter one with $g_{ij}\rightarrow t^{2/3}$ as usually found when we consider this kind of matter. The same remark is valid when the equation of state represents a radiative fluid with $\gamma=4/3$. Then, the Universe tends toward a Tolman one with $g_{ij}\rightarrow t^{1/2}$.\\
In \cite{Fay00C}, Hyperextended scalar tensor theories (HST) with a potential for the Bianchi type $I$ model are studied. HST has the same action as (\ref{1}) but with a gravitational function depending on the scalar field\cite{TorVuc96}. Some results of this last study are not changed by the presence of a perfect fluid. Especially it has been shown that the Universe isotropises when $Ge^{3\Omega}$ tends toward a constant. Applied to the present paper, it gives that $\gamma$ have to be smaller than two which is the reality condition for the equilibrium points.\\
If we consider General Relativity with only a perfect fluid, it is known that the Bianchi type $I$ model isotropises. In the present case, we observe that the presence of a scalar field add a necessary constraint, related to the range of value of the constant $\ell$, such that isotropisation might arise. We also note that the asymptotical behaviour of the metric functions does not depends on $\ell$  contrary to the case for which the scalar field dominates.\\
Last, from a mathematical point of view, the main difference between the case where the matter is absent\cite{Fay01} or does not dominate at late times and the case of trackers like theories or matter dominated theories comes from the reality condition that selects the equilibrium points. In any circumstances, the field equations written with new variables admit four equilibrium points that we will name $E_1$, $E_2$, $E_3$ and $E_4$. In the first case, reality condition selects the two first points and $\ell$ belongs to a closed interval such that $\ell^2<3$. In the second case, the two last points are selected and $\ell$ belongs to an open interval such that $\ell^2>\frac{3}{2}\gamma$.\\
\\

In what follows, we are going to study two well-known theories defined by $\sqrt{3+2\omega}\phi^{-1}=\sqrt{2}$ and some exponential and power laws potential. They have mainly been considered for FLRW models and most of the results we will get will not be new. However they will permit us to test these of the present paper and to show that the formalism we use allow unifying them. To study each of these theories, we will proceed in four steps:
\begin{enumerate}
\item We calculate the asymptotical value of the scalar field if we assume that $U\leq V^{-\gamma}$ or $U>V^{-\gamma}$.
\item We respectively deduce the conditions such that $U\leq V^{-\gamma}$ or $U>V^{-\gamma}$.
\item We respectively deduce the conditions on $\ell$ such that the Universe isotropise.
\item We compare if needed, the two sets of conditions to check their compatibility.
\end{enumerate}
\underline{The first theory} we want to study is defined by $\sqrt{3+2\omega}\phi^{-1}=\sqrt{2}$ and $U=e^{k\phi}$. In \cite{ColIbaHoo97}, it is examined without a perfect fluid and demonstrated that isotropisation arises for $k^2<2$. In \cite{Fay01} it is shown that it can not happen when $k^2>6$. If we suppose that the scalar field does not dominate at late times, we calculate that near the equilibrium $\phi$ behaves as $3\gamma k^{-1}\Omega$. Thus whatever $k$, the asymptotical behaviour of the potential is the same as this of the perfect fluid energy density and hence the solution will be trackers like solutions. The necessary condition for isotropisation related to $\ell$ is then written $k^2>3\gamma$. The calculus of $d\phi /dt$ shows that this derivative is of the same order or smaller than the potential and thus this solution is a true trackers one as usually considered in the literature. If now we assume that the scalar field asymptotically dominates the Universe and that we use the results of $\cite{Fay01}$, the asymptotical form of $\phi$ shows that our assumption is true only if $k^2<3\gamma$. The necessary condition for isotropisation given by the limit on $\ell$ is then $k^2<6$. Since $\gamma<2$, only the first inequality on $k$ have to be taken into account. Hence, when the scalar field asymptotically dominates the Universe, the necessary condition for isotropisation is satisfied. To summarise, if $k^2>3\gamma$, necessary conditions for isotropisation of the Universe toward a tracker solution such that $e^{-\Omega}\rightarrow t^{\frac{2}{3}\gamma^{-1}}$ are respected whereas if $k^2<3\gamma$, necessary conditions are respected such that it be able to isotropise toward a dominated scalar field Universe with $e^{-\Omega}\rightarrow t^{2k^{-2}}$\cite{ColIbaHoo97, Fay01}. These results have been derived in \cite{CopLidWan98} for the FLRW models. However, in this last paper, a stable trackers solution corresponding to $k^2>6$ have also been found. It does not exist in this work since then the isotropisation would be impossible.\\
\underline{The second theory} we wish to consider is defined by $\sqrt{3+2\omega}\phi^{-1}=\sqrt{2}$ and $U=\phi^k$. If we assume that the late times Universe is not asymptotically dominated by the scalar field, we find that $\phi$ will tend toward $\phi_0e^{3\gamma k^{-1}\Omega}$, $\phi_0$ being an integration constant\footnote{This constant does not appear in the previous application since it is asymptotically negligible.}. If $k<0$, $\ell$ tends to vanish and isotropisation is not possible. If $k>0$, $\ell$ diverges and necessary conditions for isotropisation are respected. Thus we deduce that the perfect fluid will asymptotically dominate this solution, hence confirming our assumption. If now we suppose that $U>V^{-\gamma}$, the calculus of the scalar field confirm it and, as shown in \cite{Fay01}, the theory is able to isotropise toward a De-Sitter model asymptotically dominated by the scalar field when $k<0$. These results are in accordance with these found in \cite{LidSch99} where it has been shown that for $k<0$, the solution is asymptotically dominated by the scalar field whereas when $k>0$, it is matter dominated.\\
\\

Some particular cases of minimally coupled scalar tensor theories with a massive scalar field and a perfect fluid has already been studied in the literature. Here we have made an attempt to derive some necessary conditions such that an asymptotically isotropic stable state be reached by the Universe at late times whatever $U$ and $\omega$ and we have then studied its dynamical behaviour. Using these results, we have made two applications and checked their consistency with previous works. In a future paper, we hope to apply the mathematical methods of this work to the Hyperextended Scalar Tensor theory for which the gravitational function varies with the scalar field.
\appendix 
\section{Divergence of $k$}\label{Ap1}
Since we have chosen to consider some positive energy densities for the perfect fluid with moreover $\gamma\in\left[1,2\right]$, then $k^2$ is positive and thus from the constraint, we deduce that the divergence of $k$ is excluded. However, in what follows, we will consider that $\delta<0$. For the constraint it is equivalent to write it as $p^2x^2+Ry^2+12z^2-k^2=1$ or to keep the same form as (\ref{11}) but with $k^2<0$. We will consider this last possibility. Then, $k$ can diverge but we wish to show that it is not compatible with an equilibrium state.\\
In a general manner, when $x\rightarrow 0$, the plane where all the isotropic stable states are present as shown in subsection \ref{s221}, we have the following relations:
\begin{equation}\label{A1}
k^2\rightarrow 1-Ry^2-12z^2
\end{equation}
and then:
\begin{equation}\label{A2}
\dot{y}\rightarrow y(6\ell z+3Ry^2-3-3/2(\gamma-2)(1-Ry^2-12z^2))
\end{equation}
\begin{equation}\label{A3}
\dot{z}\rightarrow 3Ry^2z-R/2\ell y^2-3/2(\gamma-2)z(1-Ry^2-12z^2)
\end{equation}
The expressions (\ref{A2}) and (\ref{A3}) have to tend toward zero to reach equilibrium. For the first one, it will arise if $y\rightarrow 0$ or $6\ell z+3Ry^2-3-3/2(\gamma-2)(1-Ry^2-12z^2)\rightarrow 0$. Let's study these two possibilities.\\
\\
\emph{Case 1}: $y\rightarrow 0$ and (\ref{A2})$\rightarrow 0$\\
\\
Then, (\ref{A1}) implies that $z$ diverges and (\ref{A2}) that $yz^2$ tends toward zero. Applying these two limits to (\ref{A3}), we deduce that $\dot{z}\rightarrow z^3$ and thus diverges, preventing the equilibrium. This reasoning is also valid when $\ell$ diverges.\\
\\
\emph{Case 2}: $6\ell z+3Ry^2-3-3/2(\gamma-2)(1-Ry^2-12z^2)\rightarrow 0$ and (\ref{A2})$\rightarrow 0$\\
\\
It means that:
\begin{equation}\label{A4}
y^2\rightarrow \left[-6\ell z+3+3/2(\gamma-2)(1-12z^2)\right](3/2R\gamma)^{-1}
\end{equation}
\begin{equation}\label{A5}
k^2\rightarrow 1-\left[-6\ell z+3+3/2(\gamma-2)(1-12z^2)\right](3/2\gamma)^{-1}-12z^2
\end{equation}
By putting this last expression in (\ref{A3}), we get an expression of $\dot{z}$ as a function of $z$. An equilibrium point can then be reached only for a finite value of $z$. But (\ref{A4}) shows that $y$ will tend toward a constant. Thus, it will be the same for $k$ which contradicts the fact that it becomes infinite. It is the same if $\ell$ diverges.\\
\\
Consequently, a diverging value of $k$ is not compatible with an isotropic state for the Universe at late times whatever the sign of $\delta$.


\section*{References}
\begin{thebibliography}{10}

\bibitem{Lan00}
A.~E.~Lange et~al.
\newblock First estimation of cosmological parameters from {B}oomerang.
\newblock {\em Phys.Rev.}, D63:042001, 2001.

\bibitem{Per99}
S.~Perlmutter et~al.
\newblock Measurements of ${\Omega}$ and ${\Lambda}$ from 42 {H}ight-{R}edshift
  {S}upernovae.
\newblock {\em Astrophysical Journal}, 517.

\bibitem{Rie98}
Riess et~al.
\newblock Observational evidence from {S}upernovae for an accelerating
  {U}niverse and a cosmological constant.
\newblock {\em Astrophysical Journal}, 116:1009, 1998.

\bibitem{ChiYam99}
T.~Chiba and M.~Yamaguchi.
\newblock Extended open inflation.
\newblock {\em Phys.Rev. D}, 61:027304, 1999.

\bibitem{Fay01}
S.~Fay.
\newblock Isotropisation of {G}eneralised-{S}calar {T}ensor theory plus a
  massive scalar field in the {B}ianchi type {I} model.
\newblock {\em To be published in Class. Quantum Grav}.

\bibitem{BarPie99}
N.~Bartolo and M.~Pietroni.
\newblock Scalar-{T}ensor gravity and quintessence.
\newblock {\em Phys. Rev. D}, 61:023518, 2000.

\bibitem{MatGuzUre99}
Tonatiuh Matos, Francisco~S. Guzm{\'a}n, and L.~Arturo Une{\~n}a-L{\'o}pez.
\newblock Scalar field as dark matter in the {U}niverse.
\newblock {\em Class.Quant.Grav.}, 17:1707--1712, 1999.

\bibitem{BoiEspPolSta00}
B.~Boisseau, G.~Esposito-Farese, D.~Polarski, and A.~A. Starobinsky.
\newblock Reconstruction of a scalar-tensor theory of gravity in an
  accelerating {U}niverse.
\newblock {\em Phys. Rev. Lett.}, 85:2236, 2000.

\bibitem{SerAli96}
A.~Serna and J.~M. Alimi.
\newblock Scalar-tensor cosmological models.
\newblock {\em Phys. Rev.}, D53:3074--3086, 1996.

\bibitem{SchWil01}
Paul~D. Scharre and Clifford~M. Will.
\newblock Testing scalar-tensor graviry using space gravitational wave
  interferometers.
\newblock {\em gr-qc/0109044}, page Submitted to Phys. Rev. D, 2001.

\bibitem{Nar72}
Hidekazu Nariai.
\newblock Hamiltonian approach to the dynamics of {E}xpanding {H}omogeneous
  {U}niverse in the {B}rans-{D}icke cosmology.
\newblock {\em Prog. of Theo. Phys.}, 47,6:1824, 1972.

\bibitem{MatRyaTot73}
R.~A. Matzner, M.~P. Ryan, and E.~T. Toton.
\newblock The {B}rans-{D}icke theory and anisotropic cosmologies.
\newblock {\em Nuovo Cim.}, 14B:161, 1973.

\bibitem{WaiEll97}
J.~Wainwright and G.F.R. Ellis, editors.
\newblock {\em Dynamical Systems in Cosmology}.
\newblock Cambridge University Press, 1997.

\bibitem{Fay00A}
S.~Fay.
\newblock Hamiltonian study of the generalized scalar-tensor theory with
  potential in a {B}ianchi type {I} model.
\newblock {\em Class. Quantum Grav.}, 17:891--902, 2000.

\bibitem{Mis69}
C.~W. Misner.
\newblock {\em Phys. Rev.}, 186:1319, 1969.

\bibitem{HebWet00}
A.~Hebecker and C.~Wetterich.
\newblock Quintessential adjustement of the cosmological constant.
\newblock {\em Phys.Rev.Lett.}, 85:3339--3342, 2000.

\bibitem{ChiJakPav00}
Luis~P. Chimento, Alejandro~S. Jakubi, and Diego Pav{\'o}n.
\newblock Enlarged quintessence cosmology.
\newblock {\em Phys.Rev.}, D62:063508, 2000.

\bibitem{SanKalWag98}
David~I. Santiago, Dimitri Kalligas, and Robert~V. Wagoner.
\newblock Scalar-{T}ensor {C}osmologies and their {L}ate {T}ime {E}volution.
\newblock {\em Phys. Rev.}, D58:124005, 1998.

\bibitem{Fay00C}
S.~Fay.
\newblock Exact solutions of the {H}yperextended {S}calar {T}ensor theory with
  potential in the {B}ianchi type {I} model.
\newblock {\em Class. Quantum Grav.}, 18:45, 2001.

\bibitem{TorVuc96}
Diego~F. Torres and H\'ector Vucetich.
\newblock Hyperextended scalar-tensor gravity.
\newblock {\em Phys. Rev.}, D54:7373--7377, 1996.

\bibitem{ColIbaHoo97}
A.A.Coley, J.~Ib{\`a}{\~n}ez, and R.J. van~den Hoogen.
\newblock {\em J. Math. Phys.}, 38:5256, 1997.

\bibitem{CopLidWan98}
Edmund~J. Copeland, Andrew~R. Liddle, and David Wand.
\newblock Exponential potentials and cosmological scaling solutions.
\newblock {\em Phys. Rev.}, D57:4686--4690, 1998.

\bibitem{LidSch99}
Andrew~R Liddle and Robert~J Scherrer.
\newblock A classification of scalar field potentials with cosmological scaling
  solution.
\newblock {\em Phys. Rev.}, D59:023509, 1999.

\end{thebibliography}
\end{document}